# Corrigendum

## Self-consistent theory of random lasing in the time-domain

ZACHARIAH PETERSON

Physics Department, Portland State University,
1825 SW Broadway St., Portland, OR 97201, United States

School of Business, Adams State University,
208 Edgemont Blvd., Alamosa, CO 81102, United States

*Corresponding author: petersonz1@grizzlies.adams.edu

In the online version of this article initially published (January 7, 2019), Eqs. (4) and (6) (pp. 650 and 651, respectively) were incorrect, and have been corrected as follows:

$$\begin{aligned}\dot{P} &\approx \sum_{\mu} -ik_{\mu} P_{\mu} \exp(-ik_{\mu} t) \\ &= -(ik_{\mathrm{a}} + \gamma_{\mathrm{p}}) \sum_{\mu} P_{\mu} \exp(-ik_{\mu} t) - \frac{g^2 D}{i\hbar} \sum_{\mu} \Phi_{\mu} \exp(-ik_{\mu} t)\end{aligned} \tag{4}$$

$$\nabla^2 \Phi_{\mu} + \varepsilon(\mathbf{r})(2ik_{\mu}\dot{\Phi}_{\mu} + k_{\mu}^2 \Phi_{\mu}) = \frac{4\pi i g^2}{\hbar\left[\gamma_{\mathrm{p}} + i(k_{\mathrm{a}} - k_{\mu})\right]}\left(2ik_{\mu}\frac{\partial}{\partial t}(D\Phi_{\mu}) + k_{\mu}^2 D\Phi_{\mu}\right) \tag{6}$$

*January 28, 2019*

# Self-consistent theory of random lasing in the time-domain

ZACHARIAH PETERSON

Physics Department, Portland State University,
1825 SW Broadway St., Portland, OR 97201, United States

School of Business, Adams State University,
208 Edgemont Blvd., Alamosa, CO 81102, United States

*Corresponding author: petersonz1@grizzlies.adams.edu

Random lasers are unique systems where lasing occurs due to repetitive scattering in a disordered nanostructure. Previous descriptions of random lasing are numerous, however a full time-dependent theory that describes the introduction of gain directly from first principles is lacking in the literature. This paper will present an analytic self-consistent time-dependent theory of random lasing that contains the results from the well-known steady-state *ab initio* laser theory. This theory can also describe a number of temporal phenomena that have been observed in previous experiments and facilitates the incorporation of these devices into their envisioned practical applications.

Keywords: random lasers, time-domain, condensed matter, optics.

## 1. Introduction

Since the prediction of random lasing by LETOKHOV in 1968 [1], random lasers have been extensively studied for their rich physics and potential applications as unique light sources [2, 3]. Random lasing has been observed in a number of systems, the most common of these systems being solid-state disordered semiconductor nanostructures, zinc oxide (ZnO) nanoparticles, and suspensions of gainless scatterers submersed in laser dyes [2–4]. In all of these systems, lasing arises due to repeated scattering of light within the disordered system, where the scattered light causes stimulated emission at or between successive scattering events [2, 4].

A number of classical and semiclassical theoretical models have been developed to describe laser emission from disordered media with optical gain. While conventional laser theory is based on the Maxwell–Bloch equations, early models of random lasing were based entirely on Maxwell's equations. One example is the work of BURIN *et al*. [5]; their model approximated closely packed ZnO nanoparticles as a collection of dipole oscillators. Although the model bears resemblance to the Maxwell–Bloch equations, gain was introduced phenomenologically and the model could not adequately describe the coupling between the carrier dynamics and the electromagnetic field in the disordered system. The situation was improved somewhat in [6–8]; the results in these papers are derived from an inhomogeneous wave equation model for the electric field that describes the spatial and temporal distribution of the electromagnetic field. Their model is time-dependent and has closer resemblance to the results from conventional laser theory, but the introduction of time-dependent gain was introduced phenomenologically and suffers from the same inadequacies as the model in [5].

The theory would later advance to a description of the lasing field in terms of deviations from the quasimodes of the disordered system [2, 9, 10]. DEYCH [9] calculated the deviation of a lasing mode from the eigenmodes of the system due to the introduction of optical gain in a disordered system. This work, as well as the work of TURECI *et al.* [10], would later be used to develop a steady-state solution to the Maxwell–Bloch equations in any geometry. This solution is known as steady-state *ab initio* laser theory (SALT) [11–14].

Although SALT is currently the best analytical theory to describe multimodal lasing in any geometry, it has a critical drawback that limits its application in real systems. SALT is a purely steady-state theory and it cannot be applied to systems subjected to time-dependent pumping. This drawback motivates the development of a self-consistent lasing theory that can describe the temporal dynamics of a random laser while preserving the well-known deviation of the lasing modes from the eigenmodes of the passive system.

This paper will present a fully self-consistent time-dependent theory for lasing in random media that is derived directly from the Maxwell–Bloch equations. The SALT solution and its associated results are only a subset of the theory derived in this paper. This theory has the capability to describe the temporal dynamics of the system as it approaches the SALT solution under constant pumping. The theory in this paper is not limited to constant pumping and would be immediately applicable to pulsed pump sources. As the theory is time-dependent and treats the introduction of gain directly from first principles, it also has the capability to describe the relaxation oscillations [8] and emission fluctuations [15] that have been observed in ZnO random lasers under pumping with nanosecond UV laser pulses.

## 2. The time-dependent random lasing equations

The Maxwell–Bloch equations form the basis of semiclassical laser theory and have been widely applied to conventional and random lasers; a full derivation of these equations can be found in [16]. Beginning from the Maxwell–Bloch equations for an $N$-level lasing medium, one can derive a nonlinear homogeneous wave equation that forms the cornerstone of SALT. The derivation of the SALT solution for the lasing field requires the following known inputs: the complex dielectric function $\varepsilon(\mathbf{r})$ of the cavity at the position vector $\mathbf{r}$, the polarization dephasing rate of the gain medium $\gamma_\mathrm{p}$, the relaxation rate of the inverted population $\gamma_a$, the atomic transition frequency $k_a$, and the dipole matrix element of the lasing transition $g$. The geometry of the system must also be well -defined [11–14]. SALT is also applicable in $N$-level lasing media, provided there is only a single radiative transition frequency that defines the peak of the gain spectrum in the system [14]. Certain approximations were made regarding the rate constants describing the temporal behavior of the polarization and population inversion. Specifically, $\gamma_p \sim 10^{-2}k_a$ to $10^{-3}k_a$ and $\gamma_a \sim 10^{-5}k_a$ to $10^{-7}k_a$; these values are typical for many narrowband semiconducting lasing materials [11]. These values for the rate constants also ensure that the induced polarization field closely follows the lasing field in time and the electromagnetic field is able to rise above the lasing threshold and acquire optical gain before population inversion is depleted via non-radiative relaxation processes. Under these mild assumptions, the SALT solution describes the steady state behavior in the random system and allows a full calculation of the spatial distribution of the lasing field, the lasing mode emission frequencies, their thresholds, and the emission spectrum as a function of pump strength.

In addition to the restrictions on the values of the material parameters, the SALT solution assumes the inversion always in the steady state under constant (*i.e.* time-independent) pumping. This is known within SALT formalism as the stationary inversion approximation. This approximation allows the rate equation for population inversion and the nonlinear wave equation to be decoupled, and the steady state solution of the electromagnetic field can be found by a straightforward solution algorithm. The original solution ansatz for the electromagnetic field assumed a multi-periodic time dependence with constant amplitude coefficients, *i.e.* the population inversion and electromagnetic field were in the steady state under constant pumping [11–14]. Thus the original manifestation of SALT is unable to describe the transition to the steady state solution under constant pumping. Even under constant pumping, relaxation oscillations can cause the field amplitude to fluctuate as the system transitions to the steady state [8].

SALT can be extended into the time domain by ignoring the stationary inversion approximation and allowing the inversion equation to have explicit time dependence. This allows for the treatment of random lasing systems under pulsed pumping, accounts for non-radiative relaxation of inverted population over time, and depletion of population inversion via stimulated emission. It will be shown that the time dependence of the field amplitude can be determined under the slowly-varying envelope approximation (SVEA). This approximation, along with the time-varying inversion, allows one to describe the full temporal behavior of the system. The principle results are a pair of coupled nonlinear rate equations for the electromagnetic field and population inversion that bear remarkable resemblance to the semiclassical rate equations in conventional lasers [2, 6, 7].

### 2.1. Spatial and temporal behavior of the lasing field

The analysis presented here will proceed under the same restrictions regarding material parameters that are relevant in SALT. However, we will allow the inversion and field equations to retain their explicit time-dependence. The Maxwell–Bloch equations for a 2 level system in Gaussian units are [11–13, 16]

$$\left(\frac{1}{\varepsilon(\mathbf{r})}\nabla^2 - \frac{\partial^2}{\partial t^2}\right)\mathbf{E} = \frac{4\pi}{\varepsilon(\mathbf{r})}\frac{\partial^2 \mathbf{P}}{\partial t^2} \tag{1a}$$

$$\frac{\partial \mathbf{P}}{\partial t} = -(ik_{\mathrm{a}} + \gamma_{\mathrm{p}})\mathbf{P} + \frac{g^2}{i\hbar}\mathbf{E}D \tag{1b}$$

$$\frac{\partial D}{\partial t} = \gamma_{\mathrm{a}}(D_0(\mathbf{r}, t) - D) - \frac{2}{i\hbar}(\mathbf{E}\cdot\mathbf{P}^* - \mathbf{E}^*\cdot\mathbf{P}) \tag{1c}$$

where $\mathbf{E}$ is the electric field, $\mathbf{P}$ is the polarization, $D$ is the population inversion, and $D_0(\mathbf{r}, t)$ is the pumping term. Although (1a) refers to TM modes, the extension to TE modes is elementary [11–13]. We will begin with a solution ansatz similar to SALT, and we will allow the lasing modes to have time-varying amplitudes. The series solution ansatz for the lasing modes and polarization are

$$E = \sum_{\mu} \Phi_{\mu}(\mathbf{r}, t)\exp(-ik_{\mu}t) \tag{2a}$$

$$P = \sum_{\mu} P_{\mu}(\mathbf{r}, t)\exp(-ik_{\mu}t) \tag{2b}$$

The functions $\Phi_{\mu}(\mathbf{r})$ and $P_{\mu}(\mathbf{r})$ are the spatial and temporal distribution of the electric field and polarization, respectively.

The solution to Eqs. (1a)–(1c) will proceed via substitution of (2a) and (2b) into (1b), followed by application of SVEA. The SVEA is used in conventional laser physics to reduce the derivatives of envelope functions to lower order [17]. In the following equations and in the remainder of the paper, dots will be used to denote partial derivatives with respect to time. Let $A(x, t)$ be an envelope function in space and time for the amplitude of an oscillating function $\exp(-i\omega_0 t)$; SVEA can be used to approximate the derivatives of $A(x, t)\exp(-i\omega_0 t)$ as the following:

$$\frac{\partial}{\partial t}(A\exp(-i\omega_0 t)) = (-i\omega_0 A + \dot{A})\exp(-i\omega_0 t) \approx -i\omega_0 A\exp(-i\omega_0 t) \tag{3a}$$

$$\begin{aligned}\frac{\partial^2}{\partial t^2}(A\exp(-i\omega_0 t)) &= (-\omega_0^2 A - 2i\omega_0\dot{A} + \ddot{A})\exp(-i\omega_0 t) \\ &\approx (-\omega_0^2 A - 2i\omega_0\dot{A})\exp(-i\omega_0 t)\end{aligned} \tag{3b}$$

Performing the substitution and invoking (3a) yields the following equation:

$$\begin{aligned}\dot{P} &\approx \sum_{\mu} -ik_{\mu}P_{\mu}\exp(-ik_{\mu}t) \\ &= -(ik_{\mathrm{a}} + \gamma_{\mathrm{p}})\sum_{\mu} P_{\mu}\exp(-ik_{\mu}t) - \frac{g^2 D}{i\hbar}\sum_{\mu} \Phi_{\mu}\exp(-ik_{\mu}t)\end{aligned} \tag{4}$$

Equation (4) can be solved term-by-term and the summation can be dropped. Taking the middle and right hand side of (4), the common term $\exp(-ik_\mu t)$ is canceled and we can solve for $P_\mu$:

$$P_\mu = D\Phi_\mu \frac{g^2}{i\hbar\left[\gamma_\mathrm{p} + i(k_\mathrm{a} - k_\mu)\right]} \tag{5}$$

Equation (5) can now be used in (1a) to derive the wave equation for the lasing modes. Equation (1a) can also be solved term-by-term for each value of the mode index $\mu$. Applying (1a) to each term in (2a) and (2b), cancelling the common factor $\exp(-ik_\mu t)$, and invoking (3b) for the second derivatives yields the following result:

$$\nabla^2\Phi_\mu + \varepsilon(\mathbf{r})(2ik_\mu\dot{\Phi}_\mu + k_\mu^2\Phi_\mu) = \frac{4\pi i g^2}{\hbar\left[\gamma_\mathrm{p} + i(k_\mathrm{a} - k_\mu)\right]}\left(2ik_\mu\frac{\partial}{\partial t}(D\Phi_\mu) + k_\mu^2 D\Phi_\mu\right) \tag{6}$$

In addition to applying SVEA, we have ignored products of first derivatives in (6) as these terms will evolve on the same time scale as the second derivative terms. Equation (6) can be normalized by defining an inversion scale

$$D_\mathrm{c} = \frac{\hbar\gamma_\mathrm{p}}{4\pi g^2} \tag{7}$$

and a field scale

$$E_\mathrm{c} = \frac{\hbar\sqrt{\gamma_\mathrm{p}\gamma_\mathrm{a}}}{2g} \tag{8}$$

The right hand side of (6) can be normalized using $D_\mathrm{c}$:

$$\nabla^2\Phi_\mu + \varepsilon(\mathbf{r})(2ik_\mu\dot{\Phi}_\mu + k_\mu^2\Phi_\mu) = \frac{i\gamma_\mathrm{p}}{\gamma_\mathrm{p} + i(k_\mathrm{a} - k_\mu)}\left(2ik_\mu\frac{\partial}{\partial t}(D\Phi_\mu) + k_\mu^2 D\Phi_\mu\right) \tag{9}$$

Equation (10) is the time-dependent wave equation for the lasing modes. This equation can be solved by expanding each lasing mode in the appropriate orthonormal basis [9–14] with time-dependent coefficients:

$$\Phi_\mu(\mathbf{r}, t) = \sum_m a_m^\mu(t)\varphi_m(\mathbf{r}) \tag{10}$$

The $\varphi_m(\mathbf{r})$ functions are the solutions to the Helmholtz equation for the passive cavity subject to non-Hermitian boundary conditions at the emitting interface [11–13]. These basis functions are the eigenmodes of the passive cavity and this choice of basis pre-serves the well-established connection between the passive eigenmodes and the lasing modes in disordered systems with optical gain [2, 9, 10, 18]. Let $C$ define the region of space that encompasses the disordered system, and let $\varepsilon_0$ be the dielectric function for $\mathbf{r} \notin C$; the $\varphi_m(\mathbf{r})$ functions are the solutions to

$$\nabla^2 \varphi_m(\mathbf{r}) + \varepsilon(\mathbf{r}) k_m^2 \varphi_m(\mathbf{r}) = 0, \qquad \mathbf{r} \in C \tag{11a}$$

$$\nabla^2 \varphi_m(\mathbf{r}) + \varepsilon_0 k_\mu^2 \varphi_m(\mathbf{r}) = 0, \qquad \mathbf{r} \notin C \tag{11b}$$

The non-Hermitian boundary condition at the emitting interface (also referred to as the last scattering surface $S$) is

$$(\nabla \varphi_m) \cdot \hat{n}\big|_S = i k_\mu \varphi_m\big|_S \tag{12}$$

This condition conserves photon flux emitted from the cavity [11–13, 18]. The solution is composed of purely outgoing waves under the condition

$$\lim_{r \to \infty} (\varphi_m) \propto \frac{\exp(i k_\mu r)}{r^{(n-1)/2}} \tag{13}$$

where $n$ is the dimensionality of the system. This boundary condition defines a dispersion relation between a $k_\mu$ and $k_m$, *e.g.*, $k_m \equiv k_m(k_\mu)$ [11–13]. One can show that each $k_m$ is complex with $\mathrm{Im}(k_m) < 0$ [11–13]. As the boundary condition typically results in a transcendental equation with an infinite number of solutions, one can only choose $N$ states from the entire set. A thorough discussion on the selection of states from the basis set can be found in [12, 13].

Equations (11a) and (11b) under the imposed non-Hermitian boundary condition define a Sturm–Liouville problem in $n$ dimensions. It is elementary to show that an orthonormality condition must exist in the system. The inhomogeneous dielectric function $\varepsilon(\mathbf{r})$ is a weight function for the system and defines the orthonormality condition for the basis states [19]:

$$\int \varepsilon(\mathbf{r}) \varphi_n(\mathbf{r}) \varphi_m^*(\mathbf{r}) \mathrm{d}V = \delta_{nm}, \qquad \mathbf{r} \in C \tag{14}$$

These $\varphi_m(\mathbf{r})$ functions are known as the uniform constant-flux (UCF) states within SALT [12, 13].

### 2.2. The time-dependent inversion

Now that a suitable basis expansion has been established for each mode, the inversion equation can be solved. The inversion equation in (1c) is an inhomogeneous first order

PDE that can be quickly using variation of parameters. First, the solution ansatz in (2a) and (2b) must be substituted into (1c). With the result from (5), this yields

$$\dot{D} = \gamma_a D_0(\mathbf{r}, t) - \gamma_a D\left\{1 + \frac{2g^2}{i\hbar\gamma_p\gamma_a}\left[2i\sum_{\mu}\Gamma_{\mu}|\Phi_{\mu}|^2 + \sum_{\mu\neq\nu}\frac{\Phi_{\mu}^*\Phi_{\nu}\left[i\gamma_p^2 - \gamma_p(k_a - k_{\mu})\right]}{\gamma_p^2 + (k_a - k_{\mu})^2}\exp\left[i(k_{\mu} - k_{\nu})t\right] - \text{c.c.}\right]\right\} \tag{15}$$

where

$$\Gamma_{\mu} = \frac{\gamma_p^2}{\gamma_p^2 + (k_a - k_{\mu})^2} \tag{16}$$

is the gain spectrum with width $\sim \gamma_p$. The cross terms $(\mu \neq \nu)$ in the second sum are negligible and the same result must hold under SVEA [11–13]. After normalizing the remaining sum by $E_c$ the result in (15) reduces to

$$\dot{D} = \gamma_a D_0(\mathbf{r}, t) - \gamma_a D\left(1 + \sum_{\mu}\Gamma_{\mu}|\Phi_{\mu}(\mathbf{r}, t)|^2\right) \tag{17}$$

Let

$$F(\mathbf{r}, t) = \sum_{\mu}\Gamma_{\mu}|\Phi_{\mu}(\mathbf{r}, t)|^2 \tag{18}$$

With this notation the solution to (17) is

$$D(\mathbf{r}, t) = \gamma_a \exp\left\{-\gamma_a\int_0^t\left[1 + F(\mathbf{r}, t')\right]\mathrm{d}t'\right\}\int_0^t D_0(\mathbf{r}, t')\exp\left\{\gamma_a\int_0^{t'}\left[1 + F(\mathbf{r}, s)\right]\mathrm{d}s\right\}\mathrm{d}t' \tag{19}$$

In examining Eqs. (9) and (19) we see that the inversion equation couples to the electric field nonlinearly and these equations form a pair of coupled nonlinear equations describing the full dynamics of the electric field and population inversion in space and time. We also see that each above-threshold lasing mode is coupled to the remaining above-threshold lasing modes via the $F(\mathbf{r}, t)$ term. This term embodies the nonlinear multimode interactions between spatially overlapping lasing modes and is of infinite order in general. In particular, the $F(\mathbf{r}, t)$ term also depletes the inversion and prevents divergence of the lasing solution (*i.e.* gain saturation).

### 2.3. The time-dependent field amplitude coefficients

The remaining task is to derive the equations describing the set of basis amplitudes $\{a_m^\mu\}$ for each mode. Substitution of the solution ansatz (10) into (9) yields:

$$\sum_m a_m^\mu \nabla^2 \varphi_m + \varepsilon(\mathbf{r})\varphi_m(2ik_\mu \dot{a}_m^\mu + k_\mu^2 a_m^\mu) = \frac{i\gamma_\mathrm{p}}{\gamma_\mathrm{p} + i(k_\mathrm{a} - k_\mu)} \sum_m \varphi_m \left[ 2ik_\mu \frac{\partial}{\partial t}(Da_m^\mu) + k_\mu^2 Da_m^\mu \right] \tag{20}$$

Let $\varepsilon(\mathbf{r})$ be divided into its real and imaginary components in space, *i.e.*

$$\varepsilon(\mathbf{r}) = \varepsilon_\mathrm{R}(\mathbf{r}) + i\varepsilon_\mathrm{I}(\mathbf{r}) \tag{21}$$

Gathering the differential operators with imaginary coefficients [2, 7] yields the following:

$$\sum_m \varphi_m \left[ 2\varepsilon_\mathrm{R}(\mathbf{r}) k_\mu \dot{a}_m^\mu + \varepsilon_\mathrm{I}(\mathbf{r}) k_\mu^2 a_m^\mu \right] = \sum_m \varphi_m \left[ \Gamma_\mu k_\mu^2 D a_m^\mu + \frac{2k_\mu \Gamma_\mu (k_\mathrm{a} - k_\mu)}{\gamma_\mathrm{p}} \frac{\partial}{\partial t}(Da_m^\mu) \right] \tag{22}$$

In microscopic narrowband media that we consider here, the mode spacing is large and only a small number of modes will have appreciable gain. For modes near the peak of the gain spectrum $k_\mathrm{a} - k_\mu \approx 0$. For lasing mode frequencies farther from the peak of the gain spectrum, $\Gamma_\mu \to 0$ quickly for $\gamma_\mathrm{p} \sim 10^{-2} k_\mathrm{a}$. Therefore

$$\frac{k_\mu \Gamma_\mu (k_\mathrm{a} - k_\mu)}{\gamma_\mathrm{p}} \approx 0 \;\; \forall k_\mu \tag{23}$$

and the $(\dot{D}a_m^\mu + D\dot{a}_m^\mu)$ term is negligible. Multiplying by one of the basis states $\varphi_n^*$, integrating over the entire cavity, and grouping the $\dot{a}_m^\mu$ terms together yields a system of coupled first order ODEs for $\{a_m^\mu\}$:

$$\sum_m 2\langle \varepsilon_\mathrm{R}(\mathbf{r}) \rangle_{nm} \dot{a}_m^\mu = \sum_m k_\mu \Big( \Gamma_\mu \langle D \rangle_{nm} - \langle \varepsilon_\mathrm{I}(\mathbf{r}) \rangle_{nm} \Big) a_m^\mu \tag{24}$$

In Eq. (24), $\langle g(\mathbf{r}) \rangle_{nm} = \int g(\mathbf{r}) \varphi_n^* \varphi_m \mathrm{d}V$. Equation (24) can be rewritten as a matrix equation:

$$[\varepsilon] \frac{\mathrm{d}}{\mathrm{d}t}[a] = \frac{k_\mu}{2}[M][a] \tag{25}$$

where

$$[a] = \begin{bmatrix} a_1^\mu(t) \\ \dots \\ a_m^\mu(t) \end{bmatrix} \tag{26}$$

and $[M]$, $[\varepsilon]$ are square matrices; the elements of $[M]$ are $M_{nm} = \Gamma_\mu \langle D \rangle_{nm} - \langle \varepsilon_I(\mathbf{r}) \rangle_{nm}$, and the elements of $[\varepsilon]$ are $\langle \varepsilon_R(\mathbf{r}) \rangle_{nm}$, where $n$ denotes the row number and $m$ denotes the column number. Each of the terms in $[\varepsilon]$ and $[M]$ are complex in general, thus $[a]$ may also be complex. The matrix $[a]$ contains the phase for each of the basis states and the terms collectively determine the temporal phase of a lasing mode. Because $[\varepsilon]$ and $[M]$ are both square matrices, we can solve for $\mathrm{d}[a]/\mathrm{d}t$:

$$\frac{\mathrm{d}}{\mathrm{d}t}[a] = \frac{k_\mu}{2}[\varepsilon]^{-1}[M][a] \tag{27}$$

It should be noted that Eq. (27) is valid iff $[\varepsilon]$ is non-singular. In order to guarantee this condition, the basis states should be indexed such that

$$\langle \varepsilon_R(\mathbf{r}) \rangle_{ii} > \sum_{i=1}^{N} \langle \varepsilon_R(\mathbf{r}) \rangle_{ij} \quad \forall i \tag{28}$$

Due to the fact that the $\varphi_m$ functions differ in their spatial phases and growth rates, the off-diagonal terms are likely to be smaller than the diagonal terms. Therefore, $[\varepsilon]$ is likely to have a unique inverse that can be used to invert (25) into (27). Even in the case that $[\varepsilon]$ is singular, Eq. (25) can always be used to determine $[a]$ self-consistently, albeit at greater computational expense.

Upon examining (27), we see that each $a_m^\mu(t)$ function is coupled to the remaining $a_{n\neq m}^\mu(t)$ due to the inherent disorder in the system. Equation (27) has the same form as the field amplitude equation in [2, 6, 7]. However, the field amplitude equation in [2, 6, 7] describes the introduction of gain into the system phenomenologically via rate equations for excited charge carriers and describes the effect of the dielectric function in terms of a spatial average. In contrast, Eq. (27) explicitly describes the temporal behavior for each of the lasing modes in terms of the population inversion and its coupling to the electric field. Clearly, (27) is a more complete description of the dynamic behavior of the system. Given a set of basis states $\{\varphi_m(\mathbf{r})\}$ from Eq. (11), Eqs. (19) and (27) collectively describe the full temporal dynamics of a random laser. These equations can be solved using finite difference methods by selecting the appropriate initial condition on $[a]$.

Finally, the lasing mode emission frequencies can be determined from the steady state solution using the standard SALT algorithm [11, 12]. Each real emission frequen-cy $k_\mu$ determines a set of $\{\varphi_m(\mathbf{r})\}$ that are used to compute (19) and (27). The $\{\varphi_m(\mathbf{r})\}$ basis is time-independent $\forall k_\mu$ and the coupling to the external environment via the non-Hermitian boundary condition defines the relationship between the emission frequencies and the set of complex basis state eigenvalues $\{k_m\}$. As it is well-known that the system geometry determines the allowed lasing modes in random lasers [2, 9–14, 18, 20], one can certainly use the steady state solution algorithm to determine the emission frequencies. In the UCF basis, the allowed $k_\mu$ values define a real-valued threshold; these threshold values are the solution to a threshold matrix equation (see (24) and (29) in [12]). Even above threshold, the lasing mode frequencies do not change; however, the thresholds for higher order modes may change as the pump power increases due to nonlinear multimodal interactions (*i.e.* hole-burning, gain competition, *etc.*) as the pump power increases [11–13].

### 2.4. The time-dependent power output

The power output from the system can now be derived directly from Eq. (9) under SVEA. Here will follow the same steps used in [12] to calculate the power output while paying special attention to the time-dependent terms. We will show that the equation for the power output from each mode reduces to the time-dependent analogue of Eq. (11) in [12]; the results have the same form, however the result here has a modification term due to the time-dependence of the field amplitude.

Distributing the fraction on the right-hand side of (9), invoking the narrowband approximation

$$\frac{k_\mu \Gamma_\mu (k_\mathrm{a} - k_\mu)}{\gamma_\mathrm{p}} \approx 0 \;\; \forall k_\mu \tag{29}$$

and multiplying by $\Phi_\mu^*$ yields the following:

$$\begin{aligned} &\Phi_\mu^* \nabla^2 \Phi_\mu + \varepsilon(\mathbf{r})\left(2ik_\mu \Phi_\mu^* \dot{\Phi}_\mu + k_\mu^2 |\Phi_\mu|^2\right) \\ &\quad = -2\Gamma_\mu k_\mu \left(\dot{D}|\Phi_\mu|^2 + D\Phi_\mu^* \dot{\Phi}_\mu\right) + i\Gamma_\mu k_\mu^2 D |\Phi_\mu|^2 \end{aligned} \tag{30}$$

Using the definition $\varepsilon(\mathbf{r}) = \varepsilon_\mathrm{R}(\mathbf{r}) + i\varepsilon_\mathrm{I}(\mathbf{r})$, taking (30) – (30)*, and multiplying by $1/2i$ yields the following:

$$\begin{aligned} &\frac{1}{2i}\mathrm{Im}\left(\Phi_\mu^* \nabla^2 \Phi_\mu\right) + \varepsilon(\mathbf{r}) k_\mu \frac{\mathrm{d}}{\mathrm{d}t}|\Phi_\mu|^2 + \varepsilon_\mathrm{I}(\mathbf{r}) k_\mu^2 |\Phi_\mu|^2 \\ &\quad = i\Gamma_\mu k_\mu D \frac{\mathrm{d}}{\mathrm{d}t}|\Phi_\mu|^2 + \Gamma_\mu k_\mu^2 D |\Phi_\mu|^2 \end{aligned} \tag{31}$$

The power output from the disordered system is a surface integral over the last scattering surface $S$. Invoking the non-Hermitian boundary condition and taking the surface integral over $S$ yields the result

$$\frac{k_\mu}{2}\int \nabla|\Phi_\mu|^2 \cdot \hat{n}\,\mathrm{d}S + k_\mu \frac{\mathrm{d}}{\mathrm{d}t}\int \varepsilon(\mathbf{r})|\Phi_\mu|^2 \mathrm{d}S + k_\mu^2 \int \varepsilon_\mathrm{I}(\mathbf{r})|\Phi_\mu|^2 \mathrm{d}S$$
$$= i\Gamma_\mu k_\mu \frac{\mathrm{d}}{\mathrm{d}t}\int D|\Phi_\mu|^2 \mathrm{d}S + \Gamma_\mu k_\mu^2 \int D|\Phi_\mu|^2 \mathrm{d}S \tag{32}$$

The first integral on the left-hand side of (32) is $4\pi P_\mu$, where $P_\mu$ is the power output for mode $k_\mu$ [12]. Here we have an equation for $P_\mu$ in terms of the first derivatives of the field intensity. Notice that the harmonic terms in (2a) are suppressed, and we can apply SVEA in (32) to suppress the first derivative terms proportional to $ik_\mu$. We now have the result

$$2\pi k_\mu P_\mu + k_\mu \frac{\mathrm{d}}{\mathrm{d}t}\int \varepsilon_\mathrm{R}(\mathbf{r})|\Phi_\mu|^2 \mathrm{d}S + k_\mu^2 \int \varepsilon_\mathrm{I}(\mathbf{r})|\Phi_\mu|^2 \mathrm{d}S = \Gamma_\mu k_\mu^2 \int D|\Phi_\mu|^2 \mathrm{d}S \tag{33}$$

Cancelling the common term $k_\mu$ and solving for $P_\mu$ yields the time-dependent equation for the power output:

$$P_\mu = \frac{k_\mu}{2\pi}\int \left[\Gamma_\mu D(\mathbf{r},t) - \varepsilon_\mathrm{I}(\mathbf{r})\right]|\Phi_\mu|^2 \mathrm{d}S - \frac{1}{2\pi}\frac{\mathrm{d}}{\mathrm{d}t}\int \varepsilon_\mathrm{R}(\mathbf{r})|\Phi_\mu|^2 \mathrm{d}S \tag{34}$$

### 2.5. Reduction to the SALT solution

If we take $D_0(\mathbf{r},t) \to D_0(\mathbf{r})$ and $a_m^\mu(t) \to a_m^\mu\ \forall\mu$, the integrals in Eq. (19) are trivial to evaluate. The exponential terms cancel and (19) reduces to the steady-state solution for the population inversion:

$$D_S(\mathbf{r}) = \frac{D_0(\mathbf{r})}{1 + F(\mathbf{r})} \tag{35}$$

This is identical to the steady-state inversion equation from SALT [11–13]. Taking $\dot{a}_m^\mu = 0\ \forall m,\mu$, $D(\mathbf{r},t) \to D_S(\mathbf{r})$, $\dot{D} = 0$ and reduces Eq. (9) to the steady-state wave -equation from SALT

$$\nabla^2\Phi_\mu + \varepsilon(\mathbf{r})k_\mu^2\Phi_\mu = \frac{i\gamma_\mathrm{p}}{\gamma_\mathrm{p} + i(k_\mathrm{a} - k_\mu)} k_\mu^2 D_S \Phi_\mu \tag{36}$$

Using the basis state expansion in (10) with constant coefficients and the steady-state inversion (35) in Eq. (36) yields a nonlinear homogenous wave equation for the basis functions

$$\varepsilon(\mathbf{r})\sum_m (k_\mu^2 - k_m^2)\, a_m^\mu \varphi_m = \frac{i\gamma_\mathrm{p} k_\mu^2}{\gamma_\mathrm{p} + i(k_\mathrm{a} - k_\mu)} \sum_m \frac{D_0(\mathbf{r})}{1 + F(\mathbf{r})} a_m^\mu \varphi_m \tag{37}$$

Finally, invoking the orthogonality condition for the basis states (14) we generate the threshold matrix equation that defines a system of equations for the steady-state amplitude coefficients. Multiplying (37) by $\varphi_n^*$, integrating, and solving for $a_m^\mu$ yields:

$$a_m^\mu = \frac{i\gamma_p}{\gamma_p + i(k_a - k_\mu)} \frac{k_\mu^2}{k_\mu^2 - k_m^2} \sum_n a_n^\mu \left\{ \frac{D_0(\mathbf{r})}{1 + F(\mathbf{r})} \right\}_{mn} \tag{38}$$

This is the steady-state threshold matrix equation in the UCF basis from SALT theory (Eq. (29) in [12]). The theory developed in Section 2 reduces to the well-known steady-state solution of SALT as one would expect.

## 3. Conclusions

The results in Eqs. (19), (27), and (34), taken together with the definitions in (2a) and (10), describe the full spatiotemporal dynamics of a random laser in any geometry. As in the case of SALT, this theoretical framework is self-consistent; the only required inputs are the material parameters and geometry as described in the introduction of Section 2. These equations reduce to the well-known steady-state solution of SALT. Thus the results from SALT are really a subset of the theory developed in this paper. The remaining results from SALT [12] follow logically, including the existence of nonlinear effects like mode competition, gain saturation, *etc.* [11–13]. These results provide a guide to anyone intending to design optical systems based on random lasers and obtain a dynamic description of their system.

Now that the governing equations for the full dynamic behavior have been derived, the next step is to examine the dynamic approach to the steady state solution. A forthcoming paper will use the theory developed here to analyze perturbations about the steady state using a Poincaré–Bendixson analysis, and the conditions producing stable node solutions will be examined. As will be seen in the forthcoming work, these equations can explicitly describe the relaxation oscillations in random lasers. To take this work further, the theory should be applied to a number of random systems in order to compare the numerical results with the forthcoming analytical results. The theory should also be applied to systems in the presence of spontaneous emission noise, as this is suspected to be the cause of emission fluctuations that occur in ZnO random lasers with static disorder [15].

*Acknowledgements* – This work was supported by Dr. Rolf Konenkamp's nanophotonics research group at Portland State University. The author would like to thank a useful conversation with Dr. Hui Cao and valuable guidance from Dr. Rolf Konenkamp and Dr. Robert C. Word.